\documentclass[12pt]{iopart}
\usepackage[utf8]{inputenc}
\usepackage{braket}
\usepackage{iopams}
\usepackage{setstack}
\usepackage{graphicx}
\usepackage{subcaption}
\usepackage{siunitx}
\usepackage{circuitikz}

\begin{document}

\title{Qutrit quantum battery: comparing different charging protocols}

\author{Giulia Gemme$^{1}$, Michele Grossi$^{2}$, Sofia Vallecorsa$^{2}$, Maura Sassetti$^{1,3}$ and Dario Ferraro$^{1,3}$}
\address{$^{1}$ Dipartimento di Fisica, Università di Genova, Via Dodecaneso 33, 16146 Genova, Italy\\
$^{2}$  CERN, 1 Esplanade des Particules, CH-1211 Geneva, Switzerland\\
$^{3}$  CNR-SPIN, Via Dodecaneso 33, 16146 Genova, Italy}
\ead{giulia.gemme@edu.unige.it}

\begin{abstract}
Motivated by recent experimental observations carried out in superconducting transmon circuits, we compare two different charging protocols for three-level quantum batteries based on time dependent classical pulses. In the first case the complete charging is achieved through the application of two sequential pulses, while in the second the charging occurs in a unique step applying the two pulses simultaneously. Both protocols are analytically solvable leading to a complete control on the dynamics of the quantum system. According to this it is possible to determine that the latter approach is characterized by a shorter charging time, and consequently by a greater charging power. We have then tested these protocols on IBM quantum devices based on superconducting circuits in the transmon regime. The minimum achieved charging time represents the fastest stable charging reported so far in solid state quantum batteries. 
\end{abstract}

\maketitle

\section{Introduction}
Quantum batteries (QBs) are miniaturized devices able to efficiently store and release energy on-demand exploiting the puzzling rules of quantum mechanics~\cite{Campaioli18, Bhattacharjee21}. They are intended to play a major role in the future developments of quantum technologies~\cite{Polini22}. In this direction, it is possible to imagine for example networks of QBs connected to a quantum computer with the aim of locally providing energy supply to support reversible quantum operations~\cite{Chiribella21}. The starting point of this field can be traced back to the seminal work by Alicki and Fannes in 2013~\cite{Alicki13}. Since then, the theoretical investigations have been focused on the study of the charging dynamics of one or more quantum systems each one with a finite dimension Hilbert space, usually two-level systems (TLSs)~\cite{Binder15}. 

For what it concerns the charging of QBs, two main approaches have been discussed in literature. The first one is based on the unitary energy transfer between a purely quantum charger and the QB~\cite{Andolina18, Crescente22}. This case has been discussed in particular for arrays of artificial atoms~\cite{Campaioli17, Le18, Rossini20, Rosa20, Quach20, Gyhm22} and systems for cavity and
circuit quantum electrodynamics (QED)~\cite{Ferraro18, Crescente20b, Delmonte21, Dou21, Seah21, Shaghaghi22, Zhao22, Shaghaghi23, Gemme23, Erdman22, Rodriguez23}. Remarkably enough, the first experimental evidence of a quantum charged QB has been recently reported in a system where fluorescent organic molecules play the role of two-level systems (TLSs) embedded in a microcavity~\cite{Quach22}. This system shows a behavior consistent to what predicted for the first time in~\cite{Ferraro18}, with dissipative effects counter-intuitively leading to an improvement of the stability of the QBs~\cite{Dou21}. 

On the other hand, the charging induced by a classical external drive has been also considered~\cite{Zhang19, Crescente20}. This idea culminated in the first experimental evidence of a three-level QB realized with a superconducting circuit in the transmon regime~\cite{Hu22}. The authors of this work compared two different charging protocols able to promote a qutrit (three-level quantum system) from the ground state to a second excited state. By controlling the form of the drives, they have been able to obtain both a fast and unstable and a slow and stable charging process. This latter protocol shows charging times of the order of $\approx 200\,\,\mathrm{ns}$, namely two orders of magnitude shorter with respect to the typical relaxation and dephasing times of the device ($\approx 20\,\,\mu\mathrm{s}$). In spite of this fundamental result in the field, the considered protocols were constrained by additional requirements of the field amplitudes introduced in order apply the quantum brachistochrone theory~\cite{Santos19} leading to an analytically solvable stable adiabatic charging. 

In the present paper, we will demonstrate that a full analytical solution of the qutrit's dynamics is possible under more general conditions. Reviewing the recently discussed case of a qubit QB, we will identify faster stable protocols able to realize an almost complete charging of the qutrit QB. In particular, driving the systems with properly designed Gaussian pulses, we will determine the charging time considering:

i) a sequential charging protocol where the qutrit is first promoted from the ground to the first excited state and afterwards from the first to the second excited state and 

ii) a simultaneous charging protocol where the transition directly involve the ground and the second excited state.

We will test these protocols on IBM quantum devices showing that in the simultaneous protocol the charging time can be decreased down to $\approx 20\,\,\mathrm{ns}$. 
This is an order of magnitude shorter with respect to the analysis carried out in~\cite{Hu22, Gemme22} in presence of a comparable stored energy and longer relaxation and dephasing times ($\approx 100\,\,\mu\mathrm{s}$). To the best of our knowledge this represents the fastest stable charging reported so far in the framework of QBs based on superconducting circuits, indicating the IBM quantum devices as ideal candidates to develop stable multi-level solid state quantum batteries. 

\section{Two-level QB}
\label{sec:qubit}
We start our analysis by reviewing the case of a superconducting circuit in the transmon regime 
working as a qubit (see~\cite{Koch07} and~\ref{AppA} for more details). To access quantum features these devices are put at cryogenic temperatures (few $\mathrm{mK}$). Under these working conditions, which are conventionally used in the framework of solid state quantum computation~\cite{Krantz19}, the QB can be effectively described as a two-level system with Hamiltonian (from now on we consider $\hbar=1$)
\begin{equation}
\hat{H}_{QB}^{(2)}=\omega_0\ket{0}\bra{0}+\omega_1\ket{1}\bra{1}
\end{equation}
and level spacing 
\begin{equation}
\Delta=\omega_1-\omega_0
\end{equation}
between the ground state $\ket{0}$ and the first excited state $\ket{1}$. Its dynamics is controlled by means of a classical external time-dependent drive such that the total Hamiltonian reads~\cite{Alexander20, Smith22} 
\begin{equation}
\hat{H}^{(2)}(t)=\hat{H}_{QB}^{(2)}+\hat{H}_C^{(2)}(t)   
\end{equation} 
with
\begin{equation}
\hat{H}_C^{(2)}(t)=gf(t)\cos(\Omega t)(\ket{0}\bra{1}+\ket{1}\bra{0}).
\end{equation}
In the above Equation, $f(t)$ is a time-dependent envelop function with maximum amplitude equal to one, whose form will be specified in the following. Such function is further modulated by a cosine with controllable frequency $\Omega$. Finally, $g$ represents the intensity of the (dipole) coupling between the QB and the classical drive. Notice that in our study we can safely neglect the dynamics of the external charger, due to the fact that it can be considered as a classical object not affected by the state of the QB~\cite{Crescente20}.

To study the time evolution of the state, and consequently the time behaviour of the stored energy stored, we consider the generic initial wave-function at time $t=0$~\cite{Gemme22}
\begin{equation}
\ket{\psi(0)}=\sqrt{a}\ket{0}+\sqrt{1-a}e^{i\phi}\ket{1},  
\label{Psi_general}
\end{equation}
with $0\leq a\leq 1$ and $0\leq \phi<2 \pi$ real parameters. 
The experimentally realized transmon devices, typically used in the quantum computing framework, including the devices developed by IBM, usually satisfy $g<\Delta$~\cite{Krantz19}. Under this condition, in order to achieve a complete charging of the QB, namely a perfect transition $\ket{0}\rightarrow\ket{1}$, one need to tune the frequency of the drive in such a way to precisely fulfill the condition $\Omega=\Delta$. at this point it is useful to consider the time dependent rotation
\begin{equation}
    \hat{S}^{(2)}(t)=e^{i\hat{H}_{QB}^{(2)}t}
\end{equation}
leading to the new Hamiltonian 
\begin{equation}
\hat{H}'^{(2)}=\hat{S}^{(2)}\hat{H}^{(2)}(\hat{S}^{(2)})^{\dagger}-i\hat{S}^{(2)}\frac{d(\hat{S}^{(2)})^{\dagger}}{dt}.
\label{eq:rotating_hamiltonian}
\end{equation}

Further considering the rotating wave approximation (RWA)~\cite{Schweber67, Graham84, Schleich_Book}, which is very well justified under the conditions of resonance and small coupling discussed above~\cite{Lu12}, one obtains the effective Hamiltonian
\begin{equation}
    \hat{H}^{(2)}_{eff}(t)=\frac{g}{2}f(t)(\ket{0}\bra{1}+\ket{1}\bra{0}),
\end{equation}
where we have neglected a constant term that plays no role in the dynamics. 
This leads to the Schr\"{o}dinger equation
\begin{equation}
i\ket{\dot{\psi}'(t)}=\hat{H}_{eff}^{(2)}\ket{\psi'(t)}
\end{equation}
where $\ket{\psi'(t)}=\hat{S}^{(2)}(t)\ket{\psi(t)}$ with $\ket{\psi(t)}$ the wave-function of the qubit at a given time~\cite{Gemme22}. Note that we are using the conventional Newton's dot notation to indicate the time derivative. 

Starting from this, the energy stored in the QB at the same time $t$ can be defined as~\cite{Andolina18, Ferraro18}
\begin{equation}
E^{(2)}(t)=\braket{\psi(t)|\hat{H}_{QB}^{(2)}|\psi(t)}.
\end{equation}
According to this definition and taking into account the analysis described above, the energy stored into the QB at the time $t$ can be explicitly written (assuming for now on $\omega_{0}$ as the energy reference) as 
\begin{equation}
\fl E^{(2)}(t)=\Delta  \left[a \sin^{2}{\frac{\theta(t)}{2}} +2 \sqrt{a} \sqrt{1-a} \sin\phi\sin{\frac{\theta(t)}{2}}\cos{\frac{\theta(t)}{2}}+(1-a) \cos^{2}{\frac{\theta(t)}{2}}\right],  
\label{eq:E_m_par}
\end{equation} 
where 
\begin{equation}
\theta(t)=g\int_0^{t}f(\tau)d\tau.
\end{equation} 

According to this expression the key parameter to control the system's dynamics is the area under the envelope function $f(t)$. However, in order to evaluate the energy stored into the QB as a function of time the knowledge of the form of $f(t)$ is therefore necessary. According to the analysis reported in~\cite{Gemme22}, a good choice for the envelope function is
\begin{equation}
f(t)=\mathcal{N}e^{-\frac{(t-t_m/2)^2}{2\sigma^2}},
\label{eq:Gaussian}
\end{equation} 
namely a Gaussian with amplitude $\mathcal{N}$ and standard deviation $\sigma$, centered at $t=t_{m}/2$, with $t_{m}$ the time at which the measurement of the state is carried out. In the following we will assume
\begin{equation}
\sigma=\frac{t_m}{8},
\end{equation}
where the condition $t_{m}\gg \sigma$ is fulfilled, and 
\begin{equation}
\mathcal{N}=\frac{\theta_{m}}{(g\sigma\sqrt{2\pi})},
\end{equation}
with $\theta_{m}$ the maximum amplitude achieved for the phase $\theta(t)$ induced by this pulse. 


Indeed, one has 
\begin{equation}
    \theta(t)\approx\frac{\theta_{m}}{2}\left[\text{Erf}\left(\frac{t-\frac{t_m}{2}}{\sqrt{2}\sigma }\right)+1\right],
\label{theta_t}
\end{equation}
with $\text{Erf}(x)$ the error function of argument $x$.

\begin{figure}[h]
    \centering
    \includegraphics[width=0.6\textwidth]{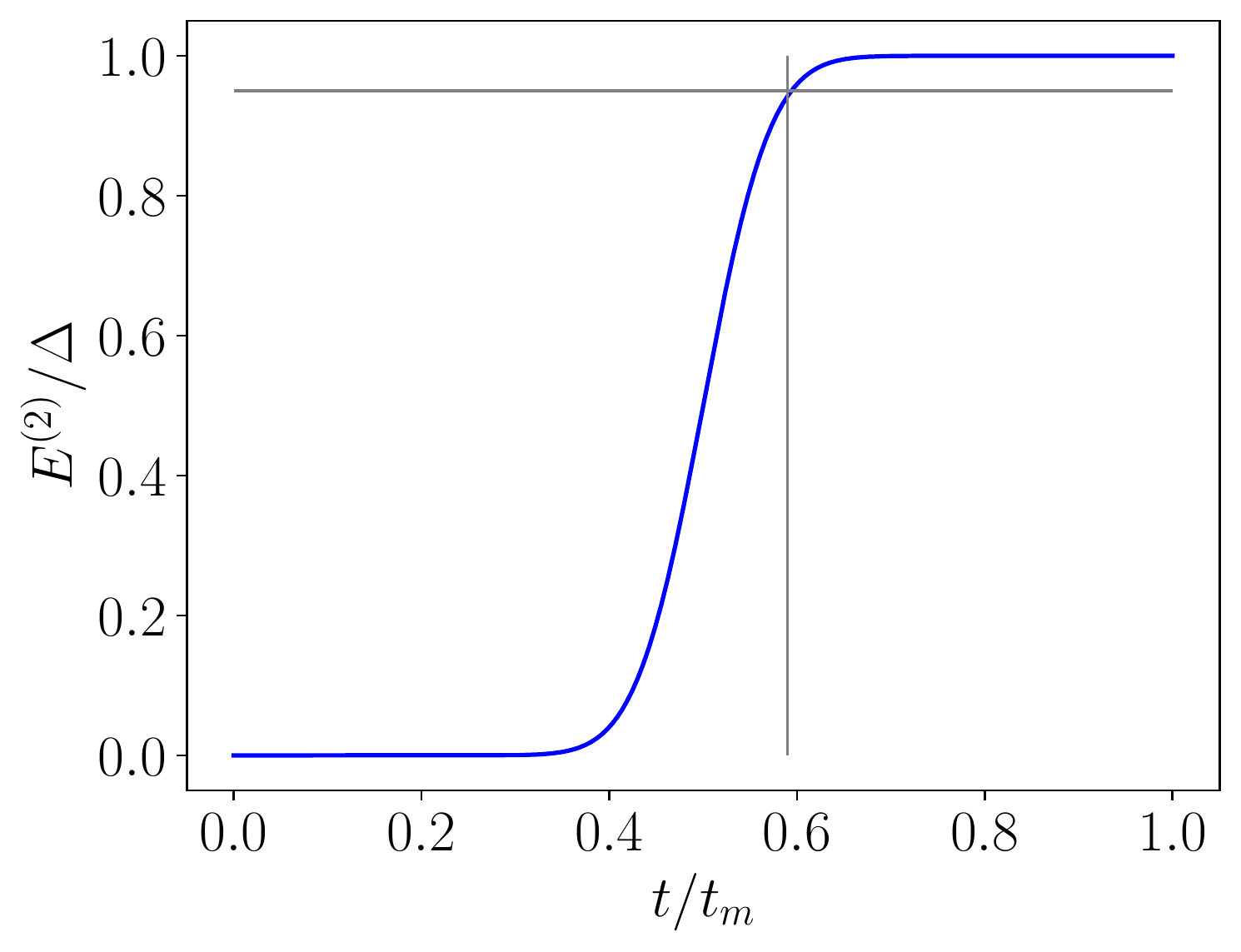}
    \caption{Blue curve: theoretical behaviour of the energy $E^{(2)}$ stored into the qubit QB (in units of $\Delta$)  as a function of $t$ (in units of $t_{m}$) and with initial condition $\ket{\psi(t)}=|0\rangle$ ($a=1$ and arbitrary $\phi$ in~(\ref{Psi_general})). Horizontal grey line indicates a QB charging of $E^{(2)}_{thr}=0.95\Delta$, while the vertical grey line is in correspondence of the charging time $t_{c}=0.59t_m$. Here we are considering $\theta_m=\pi$.}
    \label{fig:E(t)0-1}
\end{figure}

Replacing the above expression into~(\ref{eq:E_m_par}) one can determine the charging time $t_{c}$, namely the time at which the QB is (almost) completely charged, as a fraction of $t_{m}$. For example, in figure~\ref{fig:E(t)0-1}, the QB reaches a charging
$E^{(2)}_{thr}=0.95\Delta$ for a time $t_{c}=0.59t_m$.

\begin{figure}[h]
	\centering
	\begin{subfigure}{0.49\textwidth}
		\centering
        \includegraphics[width=1\textwidth]{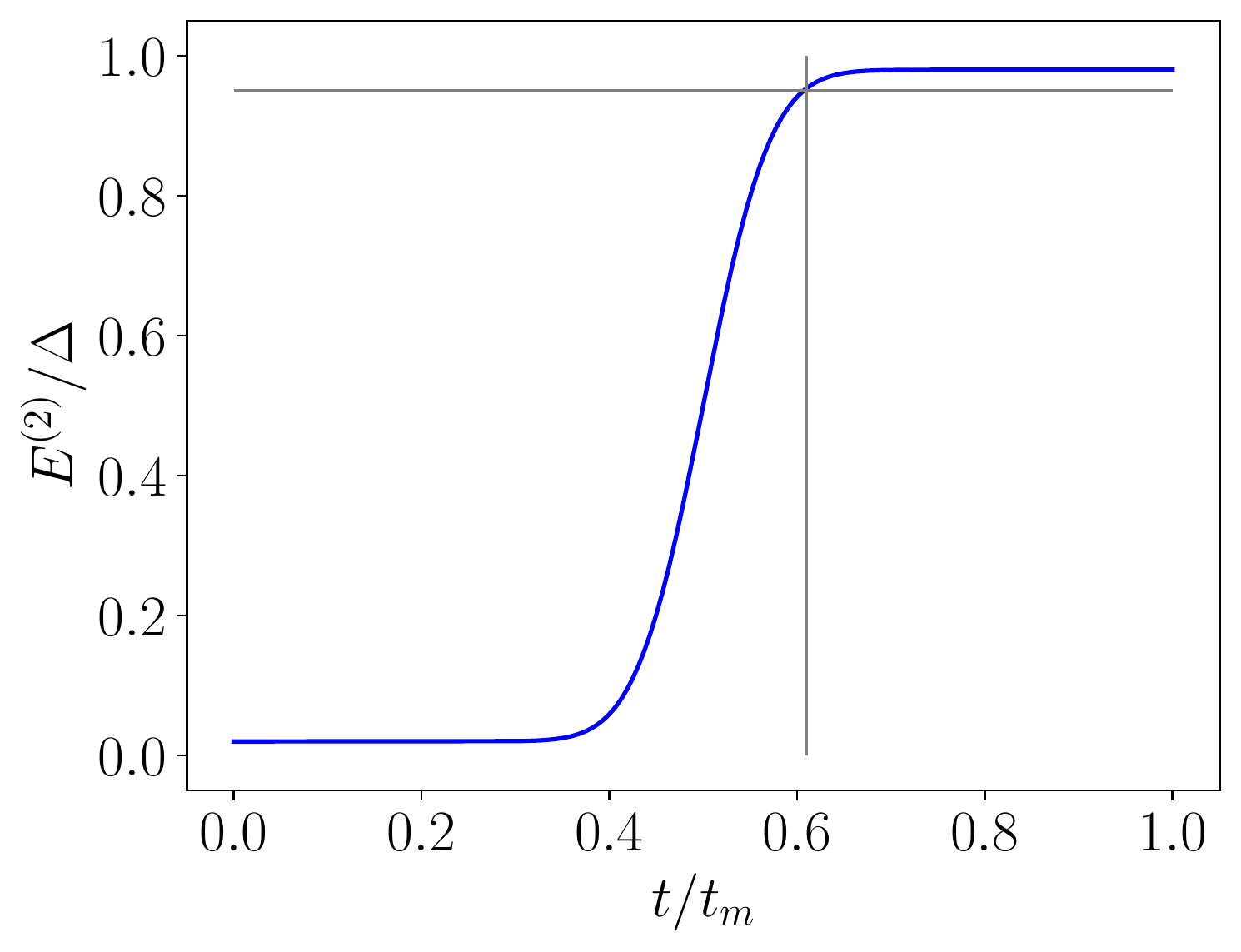}
        \caption{}
        \label{fig:E(t)0-1_098}
	\end{subfigure}
	\begin{subfigure}{0.49\textwidth}
		\centering
		\includegraphics[width=1\textwidth]{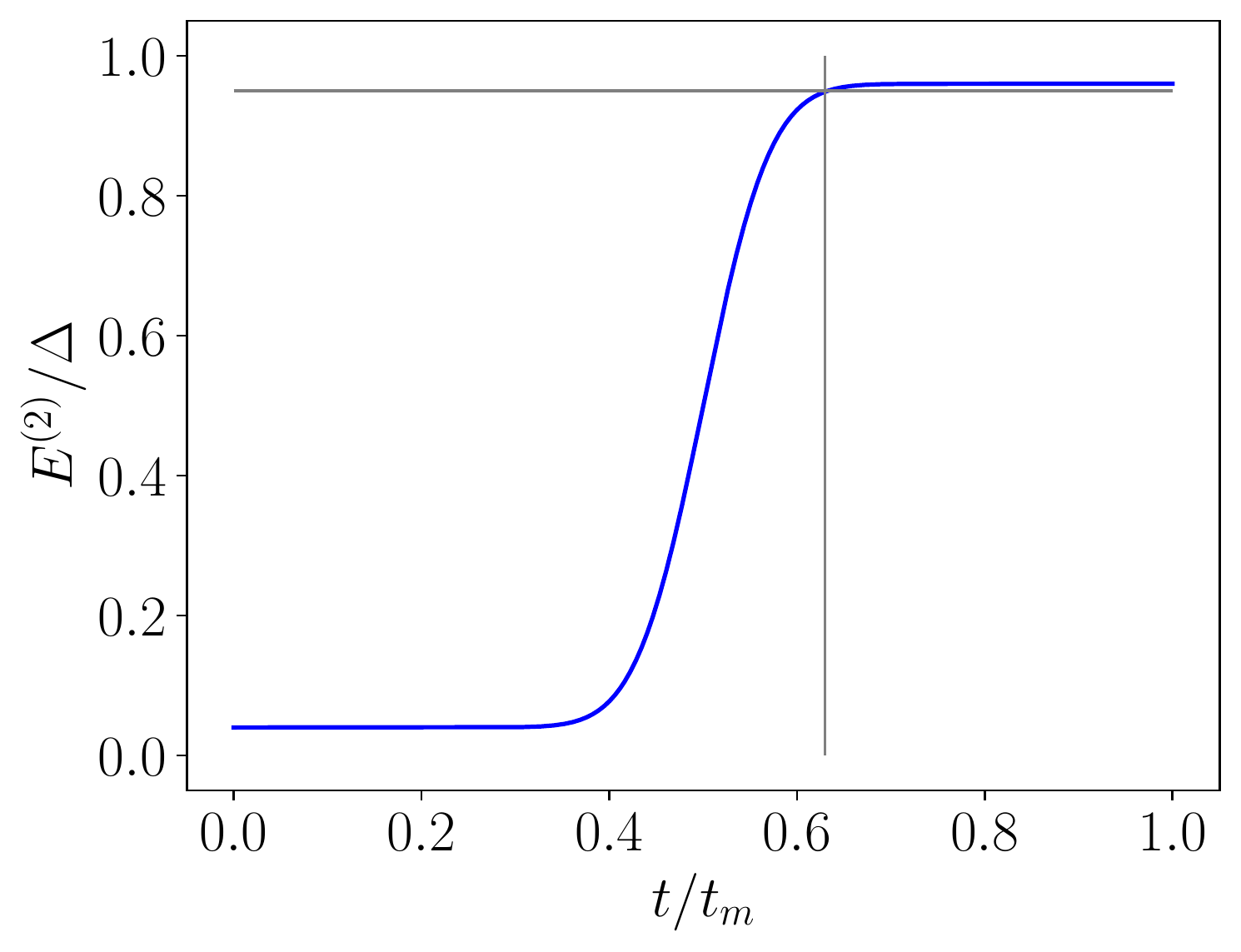}
		\caption{}
		\label{fig:E(t)0-1_096}
	\end{subfigure}
	\caption{Blue curves: theoretical behaviour of the energy $E^{(2)}$ stored into the qubit QB (in units of $\Delta$) as a function of $t$ (in units of $t_{m}$) and with $a=0.98$, $\phi=0$ (a) and $a=0.96$, $\phi=0$ (b) respectively. We have considered $E^{(2)}_{thr}=0.95\Delta$ in both panels (horizontal grey lines). This leads to $t_{c}=0.61 t_m$ (a) and $t_{c}=0.63 t_m$ (b) respectively (vertical grey lines). Here we are considering $\theta_m=\pi$.}
\end{figure}

It is also useful to consider more realistic situations. Indeed, according to the analysis reported in~\cite{Gemme22}, in a real device it is not possible to initialize the system exactly in the ground state. According to this, considering for example the conditions $a=0.98$, $\phi=0$ (figure~\ref{fig:E(t)0-1_098}) and $a=0.96$, $\phi=0$ (figure~\ref{fig:E(t)0-1_096}), the charging $E^{(2)}_{thr}=0.95\Delta$ is achieved for $t_{c}=0.61 t_m$ and $t_{c}=0.63 t_m$ respectively.

Obviously, the arbitrary choice of the value for the threshold $E^{(2)}_{thr}$ could play in general a relevant role in determining the charging time. However, assuming $E^{(2)}_{thr}$ ranging form $0.92\Delta$ to $0.99\Delta$ the charging times are only marginally different, namely $t_{c}\approx 0.6 t_m$ (see table~\ref{label}). This strengthens the validity of our estimation.  
\begin{table}
\caption{\label{label}Charging times (in units of $t_{m}$) for different initial states of the QB (denoted by $a$ and $\phi$ according to~(\ref{Psi_general}) and values of the energy threshold.}
\begin{indented}
\item[]\begin{tabular}{@{}llll}
\br
$a$ & $\phi$ & $E_{thr}/\Delta$ & $t_c/t_m$\\
\mr
1&0&0.92&0.58\\
1&0&0.95&0.59\\
1&0&0.99&0.63\\
0.98&0&0.95&0.61\\
0.98&$\frac{\pi}{4}$&0.95&0.63\\
0.96&0&0.95&0.63\\
0.96&$\frac{\pi}{4}$&0.95&0.68\\
\br
\end{tabular}
\end{indented}
\end{table}

\begin{figure}[h]
    \centering    \includegraphics[width=0.6\textwidth]{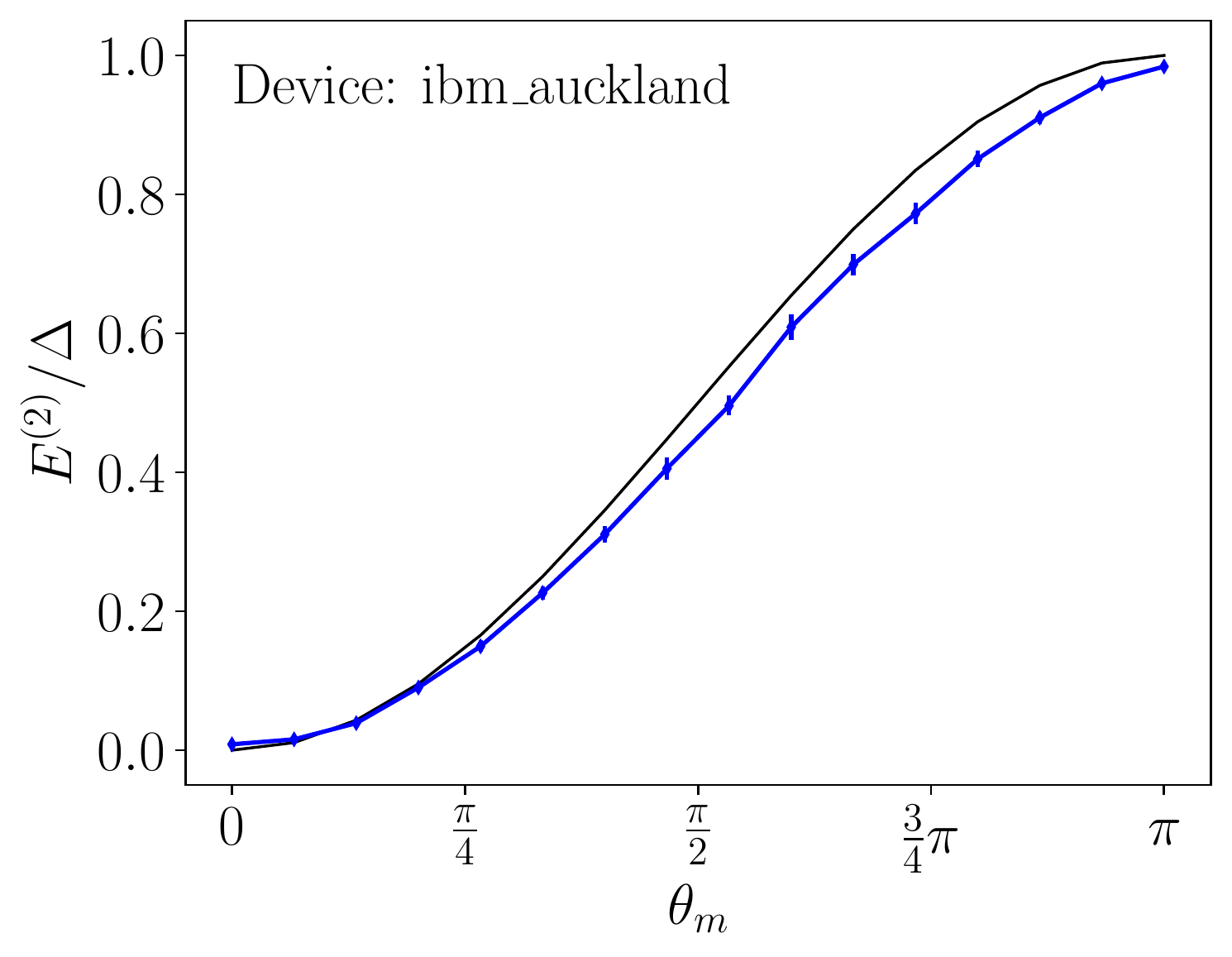}
    \caption{Energy stored in the QB (in units of $\Delta$) as a function of $\theta_{m}$. The black line is obtained analytically using~(\ref{eq:E_m_par}) and with initial condition $\ket{\psi(t)}=|0\rangle$ ($a=1$ and arbitrary $\phi$ in~(\ref{Psi_general})). The blue points correspond to experimental data obtained from the \emph{ibm\_auckland} device, using the Gaussian pulses described in the main text with $t_m=\SI{30}{ns}$.}
    \label{fig:E(theta)0-1E95}
\end{figure}

Despite the above analysis, the charging behaviour in real time cannot be directly addressed in a cloud based access as the one provided by IBM. However, it is possible to reconstruct it starting from the evolution of the stored energy at a fixed measurement time $t=t_m$ and as function of $\theta_{m}$. The theoretical prediction for this quantity in the case of an ideal charging starting from the ground state $|0\rangle$ is reported in figure~\ref{fig:E(theta)0-1E95} and compared to the real data extracted from the IBM quantum machine $\emph{ibm\_auckland}$. Data are extracted from the machine following the calibration procedure described in~\cite{Gemme22}. Notice that this curve does not depends on the functional form of the drive, provided that $t_{m}\gg \sigma$ with $\sigma$ the typical width associated to $f(t)$, and that the deviation with respect to the theoretical prediction mainly depends on the fact that the system cannot be initialized exactly in the ground state, that the pulses are discretized and to possible read-out errors~\cite{Gemme22}. Taking into account the fact that for this specific experiment $t_{m}=\SI{30}{ns}$ and due to the above considerations, one can estimate a charging time $t_{c}\approx\SI{20}{ns}$. This value is orders of magnitude shorter with respect to the decay time of the device ($\approx \SI{100}{\mu s}$), leading to a great stability of the QB~\cite{Carrega20}. Moreover, this time is shorter with respect to the one achieved in~\cite{Gemme22} due to the greater values of coupling characterizing $\emph{ibm\_auckland}$ in comparison with the one of $\emph{ibm\_armonk}$ used there ($g \approx 1 \text{ Ghz}$ vs $g \approx 0.1 \text{ Ghz}$).  

\section{Three-level QB}
\label{sec:qutrit}
We want now to investigate the possibility to realize charging protocols addressing the ground ($|0\rangle$) and first two excited states ($|1\rangle$, $|2\rangle$) of a transmon, namely realizing a qutrit QB described by the Hamiltonian (see \ref{AppA} for more details)
\begin{equation}
\hat{H}_{QB}^{(3)}=\omega_0\ket{0}\bra{0}+\omega_1\ket{1}\bra{1}+\omega_2\ket{2}\bra{2}.
\end{equation}
Also in this case the dynamics is controlled by means of classical external drives such that
\begin{equation}
\hat{H}^{(3)}(t)=\hat{H}_{QB}^{(3)}+\hat{H}_C^{(3)}(t)
\end{equation}
with
\begin{eqnarray}
\hat{H}_C^{(3)}(t)&=&gf_1(t)\cos(\Omega_1 t)(\ket{0}\bra{1}+\ket{1}\bra{0})\nonumber\\
&+&gf_2(t)\cos(\Omega_2 t)(\ket{1}\bra{2}+\ket{2}\bra{1}). 
\label{Charging_3}
\end{eqnarray}
Here, $f_{1}(t)$ and $f_{2}(t)$ are two generally different time dependent envelope functions generalizing what shown in the previous Section.

Notice that the general form of the classical driving Hamiltonian in~(\ref{Charging_3}) allows for a direct coupling only between states with opposite parity ($|0\rangle \leftrightarrow |1\rangle$, $|1\rangle \leftrightarrow |2\rangle$). Despite the fact that a transition $|0\rangle \leftrightarrow |2\rangle$ could be useful for the following analysis, it cannot be implemented in these machines. The Hamiltonian in (\ref{Charging_3}) has been also investigated in~\cite{Hu22}. However, in the following we will discuss more versatile and efficient charging protocols, leading to a faster and stable charging. 

Proceeding in full analogy with what done in the case of the qubit QB one can consider a time dependent rotation of the form
\begin{equation}
\hat{S}^{(3)}(t)=e^{i\hat{H}_{QB}^{(3)}t}
\end{equation}
to describe the system in the rotating frame. 
In order to simplify the notation one can define:
\begin{eqnarray}
    \Delta&=&\omega_1-\omega_0, \\ \Delta'&=&\omega_2-\omega_1.
\end{eqnarray}
In the transmon geometry considered in this work, one has $\Delta>\Delta'$ due to the fact that this device can be described as a anharmonic oscillator of the Duffing type (see~\cite{Koch07, Krantz19} and \ref{AppA} for more details). 

Considering again the RWA valid for $\Omega_1 = \Delta$ and $\Omega_2 = \Delta'$, the effective Hamiltonian in the rotating frame can be evaluate by means of the relation 
\begin{equation}
\hat{H}'^{(3)}(t)=\hat{S}^{(3)}\hat{H}^{(3)}(\hat{S}^{(3)})^{\dagger}-i\hat{S}^{(3)}\frac{d(\hat{S}^{(3)})^{\dagger}}{dt}
\label{eq:rotating_hamiltonian3}
\end{equation} 
and reads
\begin{equation}
\hat{H}_{eff}^{(3)}=\frac{g}{2}f_1(t)(\ket{0}\bra{1}+\ket{1}\bra{0})+\frac{g}{2}f_2(t)(\ket{1}\bra{2}+\ket{2}\bra{1}).
\end{equation}
This leads to the Schr\"{o}dinger equation
\begin{equation}
i\ket{\dot{\Psi}'(t)}=\hat{H}_{eff}^{(3)}\ket{\Psi'(t)}
\end{equation}
where $\ket{\Psi'(t)}=\hat{S}^{(3)}(t)\ket{\Psi(t)}$ with $\ket{\Psi(t)}$ the wave-function of the qutrit at a given time.

Considering the conventional spinorial notation 
\begin{equation}
    \ket{\Psi'(t)}=
    \left( \matrix{
        c_2(t)\cr
        c_1(t)\cr
        c_0(t)
    \cr} \right)
\end{equation}
the dynamics of the system is obtained by solving the set of differential equations
\begin{equation}
    \left( \matrix{
        \dot{c}_2(t)\cr
        \dot{c}_1(t)\cr
        \dot{c}_0(t)
    \cr} \right)=
    -i\frac{g}{2}\left( \matrix{
        0 & f_2(t) & 0\cr
        f_2(t) & 0 & f_1(t)\cr
        0 & f_1(t) & 0
    \cr} \right)\left( \matrix{
        c_2(t)\cr
        c_1(t)\cr
        c_0(t)
    \cr} \right).
\end{equation}
According to the previous discussion, in order to excite the QB from $|0\rangle$ to $|2\rangle$ we have to apply two pulses to the system. In order to achieve this goal in the two following Subsections we will address two different situations: i) the two pulses are applied sequentially and ii) two pulses are simultaneous. As will be clear in the following, both these cases can be treated analytically. From a theoretical point of view also intermediate situations where the two pulses partially overlap can be studied by considering a numerical approach, however we are not going to discuss these cases in details because it not easy to directly implement them on IBM machines~\cite{IBM}.

In the following, we will evaluate the energy stored in the three-level QB
\begin{equation}
   E^{(3)}(t)=\braket{\Psi(t)|\hat{H}_{QB}^{(3)}|\Psi(t)}\\
\end{equation}
assuming the ground state  of the qutrit as initial state, namely
\begin{equation}
\ket{\Psi(0)}=\ket{0}.    
\end{equation}
We will also comment about possible deviations with respect to this ideal condition in realistic implementations. Notice that, despite the different physical implementation and objectives,  the formalism we are considering presents analogies with recently reported protocols for coherent energy transfer~\cite{Crescente23}. 

\subsection{Sequential charging protocol}
\label{subsec:two-step}
Here, we can choose two identical, but properly delayed in time, pulses, namely $f_2(t)=f_1(t-t_m/2)$ and with $f_1(t)$ of the same Gaussian form as in~(\ref{eq:Gaussian}) but with $t_m\to t_m/2$, in such a way that the total duration of the protocol is $t_m$. In this limit, one can analytically solve the problem in two steps, each one identical to what previously discussed in the case of the qubit.

\subsubsection*{$\ket{0}\to\ket{1}$ transition.}

In this case one needs to solve the set of differential equations:
\begin{equation}
    \left( \matrix{
        \dot{c}_2(t)\cr
        \dot{c}_1(t)\cr
        \dot{c}_0(t)
    \cr} \right)=
    -i\frac{g}{2}\left( \matrix{
        0 & 0 & 0\cr
        0 & 0 & f_1(t)\cr
        0 & f_1(t) & 0
    \cr} \right)\left( \matrix{
        c_2(t)\cr
        c_1(t)\cr
        c_0(t)
    \cr} \right).
\end{equation}
The energy stored in the QB in this phase is (see (\ref{eq:E_m_par}) with $a=1$)
\begin{equation}
   E^{(3)}_{seq}(t)=\Delta\sin^2\frac{\theta_1(t)}{2}
\end{equation}
with 
\begin{equation}
\theta_1(t)=g\int_0^tf_1(\tau)d\tau
\end{equation}
and $t\in[0,t_m/2]$. Notice that one can safely assume that out of this interval $f_{1}(t)$ is essentially zero. 

\subsubsection*{$\ket{1}\to\ket{2}$ transition.}
Here, we need to solve the set of differential equations
\begin{equation}
    \left( \matrix{
        \dot{c}_2(t)\cr
        \dot{c}_1(t)\cr
        \dot{c}_0(t)
    \cr} \right)=
    -i\frac{g}{2}\left( \matrix{
        0 & f_2(t) & 0\cr
        f_2(t) & 0 & 0\cr
        0 & 0 & 0
    \cr} \right)\left( \matrix{
        c_2(t)\cr
        c_1(t)\cr
        c_0(t)
    \cr} \right).
\end{equation}
Assuming that in the previous step the system reaches the first excited state ($|\Psi(t_{m}/2)\rangle\approx|1\rangle$), the energy stored in the QB is given by
\begin{equation}
E^{(3)}_{seq}(t)=\Delta+\Delta'\sin^2\frac{\theta_2(t)}{2}
\end{equation}
with 
\begin{equation}
\theta_2(t)=g\int_{\frac{t_m}{2}}^{t}f_2(\tau)d\tau
\end{equation}
and $t\in[t_m/2,t_m]$. Also in this case, out of this interval $f_{2}(t)$ can be considered as null. Using the same Gaussian envelope function discussed in previous Section, with $\sigma=t_m/16$, one has

\begin{eqnarray}
    \theta_{1}(t)&\approx&\frac{\theta_{1, m}}{2}\left[\text{Erf}\left(\frac{t-\frac{t_m}{4}}{\sqrt{2}\sigma }\right)+1\right]\\
    \theta_{2}(t)&\approx&\frac{\theta_{2,m}}{2}\left[\text{Erf}\left(\frac{t-\frac{3t_m}{4}}{\sqrt{2}\sigma }\right)+1\right].
\label{varphi_t}
\end{eqnarray}

\begin{figure}[h]
    \centering
    \includegraphics[width=0.6\textwidth]{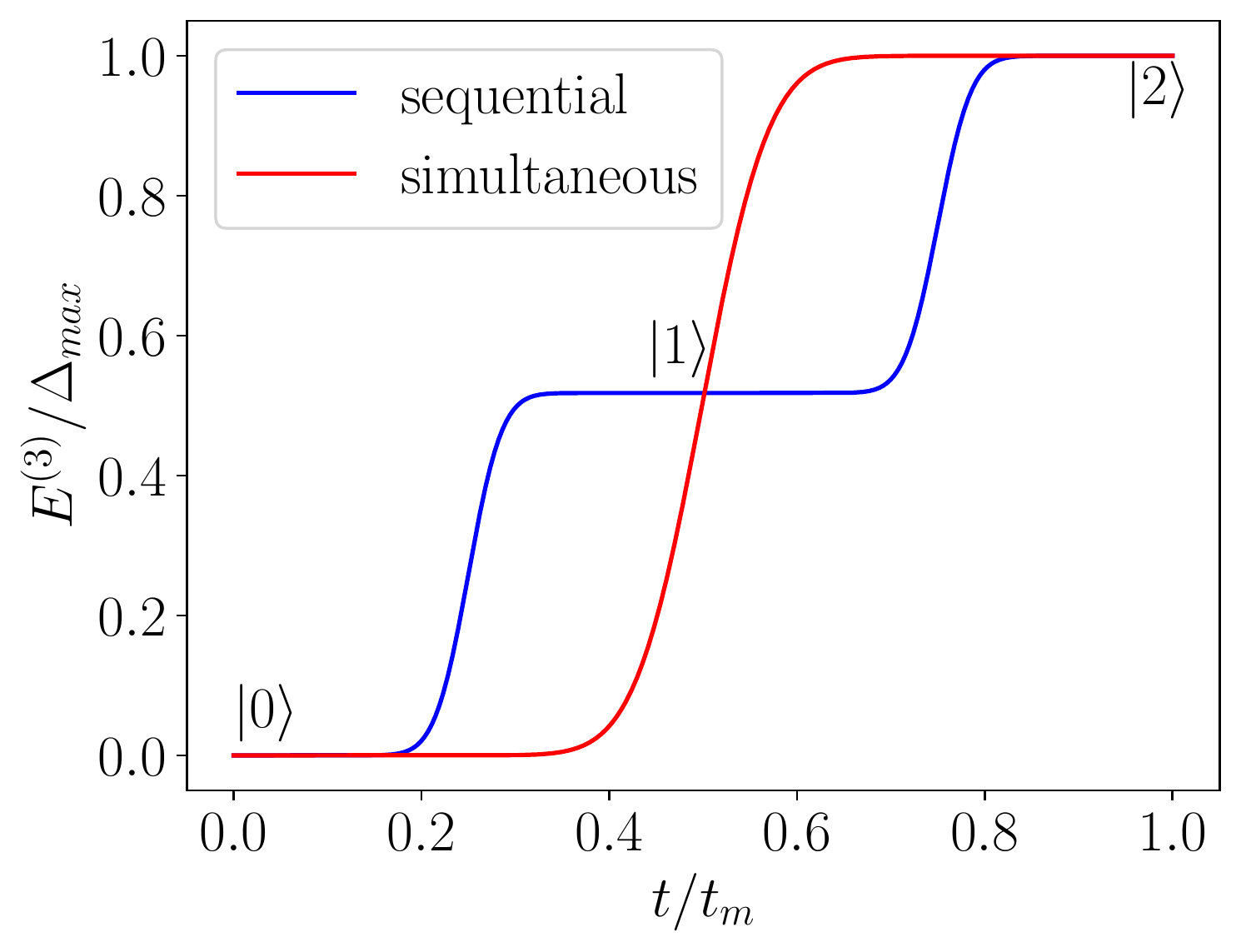}
\caption{Energy stored in the QB (in units of $\Delta_{max}=\Delta+\Delta'$) as a function of $t$ (in units of $t_{m}$) for both a sequential (blue curve) and simultaneous (red curve) charging protocol. Here we have considered pulses of Gaussian form satisfying the constrains discussed in the main text, with $\varphi_m=2\pi$, $\Theta_m=\pi$ and $a=1$.}
    \label{fig:alignVSsequential_time}
\end{figure}

In figure~\ref{fig:alignVSsequential_time} we show the energy stored in the battery as a function of time. Here, one can clearly see a two-step charging (blue curve). For what it concerns the charging time, due to the similarity with the qubit charging the same estimation discussed above works also here, limited to the second step ($t_c \approx 0.8 t_m$).

Also in this case the real time dynamics cannot be accessed directly in IBM quantum devices. It is however possible to extract it from the behaviour of the energy stored at the final measurement time $t_{m}$ and  as a function of the quantity 
\begin{equation}
    \varphi_{m}=\left\{\begin{array}{l@{\quad}cr} 
        \theta_{1,m} & \mathrm{if}  & \theta_{1,m}\in[0,\pi]\\
        \pi + \theta_{2,m} & \mathrm{if}  & \theta_{1,m}=\pi \, \& \, \theta_{2,m}\in[0,\pi].
 \end{array}\right.
    \label{eq:energy-two_step}
\end{equation}
In terms of this new variable one has
\begin{equation}
    E^{(3)}_{seq}(\varphi_{m})=\left\{\begin{array}{l@{\quad}cr} 
        \Delta\sin^2\frac{\varphi_{m}}{2} & \mathrm{if}  & \varphi_{m}\in[0,\pi]\\
        \Delta+\Delta'\sin^2\left(\frac{\varphi_{m}-\pi}{2}\right) & \mathrm{if}  & \varphi_{m}\in[\pi,2\pi].
 \end{array}\right.
    \label{eq:energy-two_step}
\end{equation}

\begin{figure}[h]
    \centering
    \includegraphics[width=0.6\textwidth]{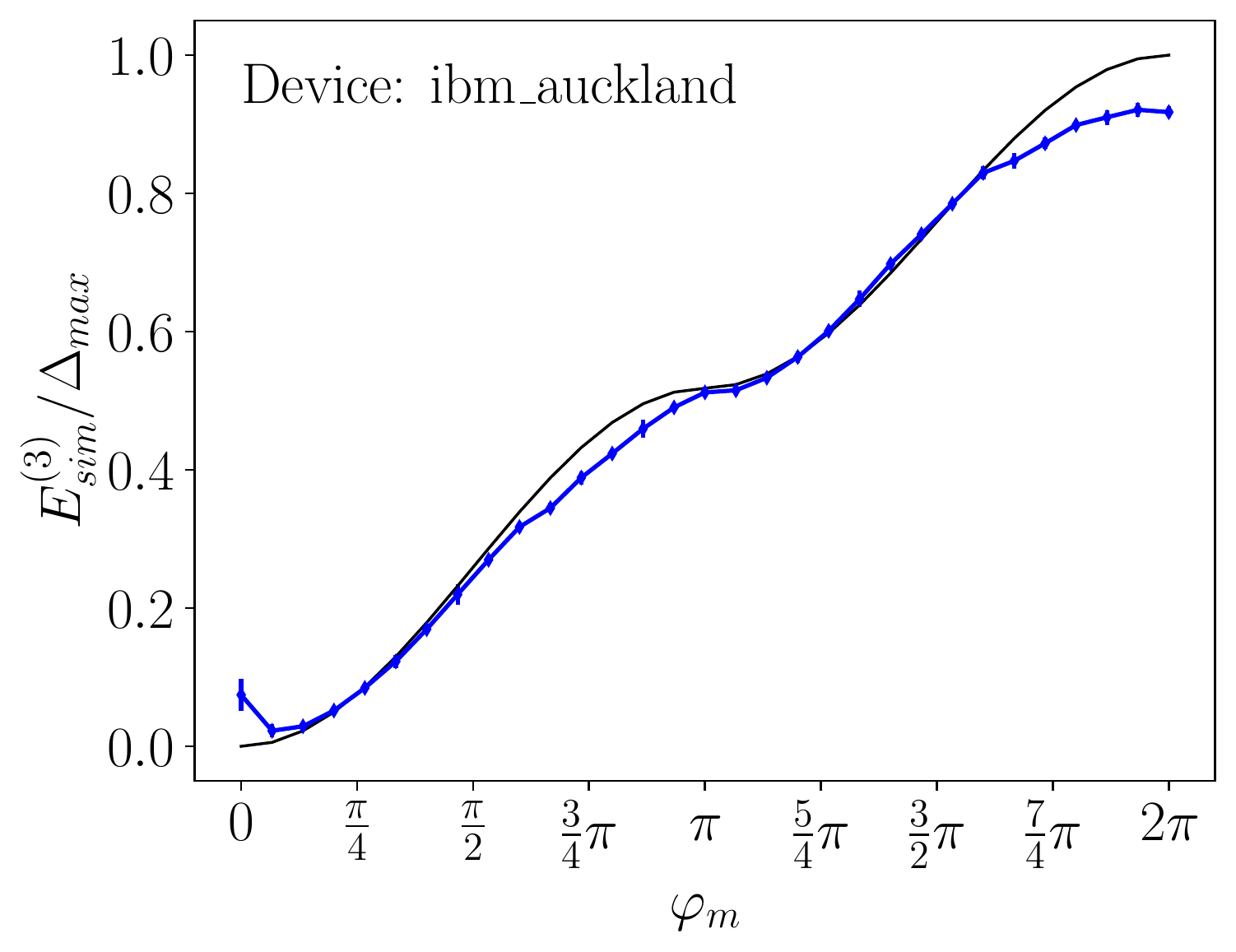}
    \caption{Energy stored in the QB (in units of $\Delta_{max}$) as a function of $\varphi_{m}$ following the sequential charging protocol. The black line is obtained analytically using~(\ref{eq:energy-two_step}). The blue points correspond to experimental data obtained from the \emph{ibm\_auckland} device, using the Gaussian pulses described in the main text with $t_m=\SI{30}{ns}$.}
    \label{fig:seq_medie15ns}
\end{figure}

The behaviour of the above function, together with the the relative experimental data obtained using the $\emph{ibm\_auckland}$ device (a machine composed by $27$ transmon circuits, of which we address the number $0$ that is characterized by the the best compromise between the longer relaxation and dephasing times and the smaller read-out error), are reported in figure~\ref{fig:seq_medie15ns}. Data are extracted from the machine following the calibration procedure described in~\ref{sec:calibration}. The agreement between data and the theoretical function is very good, in particular in the first half of each step. However, the experimental data show that it is not possible to fully charge the QB. Indeed, the maximum energy reached is $92.1\%$ of the maximum energy $\Delta_{max}=\Delta+\Delta'$.

Here, the charging occurs in a time ($t_{c}\approx 25\,\,\mathrm{ns}$) which is way shorted with respect to the decay time of the device ($\approx 100\,\,\mu\mathrm{s}$). This stable charging protocol is similar to the one discussed in~\cite{Hu22}, although there the charging was reached in the longer time $t_{c}\approx200\,\,\mathrm{ns}$. This faster charging is a consequence of the stronger dipole coupling characterizing this quantum device. For what it concerns the amount of energy stored in the qutrit, one has $\Delta_{max}\approx 39.2\,\,\mu\mathrm{eV}$, which is smaller but of the same order of magnitude of the one reported in~\cite{Hu22} ($\Delta_{max}\approx 50.6\,\,\mu\mathrm{eV}$).  

\subsection{Simultaneous charging protocol}
\label{subsec:one-step}
In this case one has $f_1(t)=f_2(t)=f(t)$, which leads to the set of differential equation
\begin{equation}
    \left( \matrix{
        \dot{c}_2(t)\cr
        \dot{c}_1(t)\cr
        \dot{c}_0(t)
    \cr} \right)=
    -i\frac{g}{2}f(t)\left( \matrix{
        0 & 1 & 0\cr
        1 & 0 & 1\cr
        0 & 1 & 0
    \cr} \right)\left( \matrix{
        c_2(t)\cr
        c_1(t)\cr
        c_0(t)
    \cr} \right).
\end{equation}
The matrix 
\begin{equation}
\hat{T}=\left( \matrix{
        0 & 1 & 0\cr
        1 & 0 & 1\cr
        0 & 1 & 0
    \cr} \right)
\end{equation}
is diagonalized by means of the unitary transformation

\begin{equation}
\hat{\mathcal{U}}=\left( \matrix{
        \frac{1}{2} & -\frac{1}{\sqrt{2}} & \frac{1}{2}\cr
        \frac{1}{2} & \frac{1}{\sqrt{2}} & \frac{1}{2}\cr
        \frac{1}{\sqrt{2}} & 0 & \frac{1}{\sqrt{2}}
    \cr} \right).
\end{equation}
This leads to new set of the decoupled equations
\begin{equation}\fl
    \left( \matrix{
        \dot{c}_-(t)\cr
        \dot{c}_+(t)\cr
        \dot{c}_B(t)
    \cr} \right)=
    -i\frac{g}{2}f(t)\left( \matrix{
        -\sqrt{2}c_-(t)\cr
        \sqrt{2}c_+(t)\cr
        0
    \cr} \right) \quad \to \quad \left\{  \begin{array}{l@{\quad}cr} 
    c_-(t)=e^{i\frac{g}{\sqrt{2}}\int_0^tf(\tau)d\tau}c_-(0)\\
    c_+(t)=e^{-i\frac{g}{\sqrt{2}}\int_0^tf(\tau)d\tau}c_+(0)\\
    c_B(t)=c_B(0).
     \end{array}\right.
\end{equation}
In this basis the state at time $t$ is given by  
\begin{equation}
    \ket{\Psi'(t)}=\left( \matrix{
    e^{i\Theta(t)}c_-(0)\cr
    e^{-i\Theta(t)}c_+(0)\cr
    c_B(0)
    \cr} \right)
\end{equation}
with 
\begin{equation}
\Theta(t)=\frac{g}{\sqrt{2}}\int_0^tf(\tau)d\tau.
\end{equation}
Is this new basis the initial conditions lead to
\begin{equation}
    \left( \matrix{
        c_-(0)\cr
        c_+(0)\cr
        c_B(0)
    \cr} \right)=\left( \matrix{
        \frac{1}{2}\cr
        \frac{1}{2}\cr
        \frac{1}{\sqrt{2}}
    \cr} \right)
\end{equation}
and consequently
\begin{equation}
    \ket{\Psi'(t)}=\left( \matrix{
    \frac{1}{2}e^{i\Theta(t)}\cr
    \frac{1}{2}e^{-i\Theta(t)}\cr
    \frac{1}{\sqrt{2}}
    \cr} \right).
\end{equation}
Returning back to the original basis we finally have
\begin{equation}
    \ket{\Psi'(t)}=\left( \matrix{
    \frac{1}{2}\left[\cos\Theta(t)-1\right]\cr
    -\frac{i}{\sqrt{2}}\sin\Theta(t)\cr
    \frac{1}{2}\left[\cos\Theta(t)+1\right]
    \cr}\right).
\end{equation}
Considering the same Gaussian pulse as in the qubit case one obtains 
\begin{equation}
    \Theta(t)=\frac{\Theta_{m}}{2}\left[\text{Erf}\left(\frac{t-\frac{t_m}{2}}{\sqrt{2}\sigma }\right)+1\right]
\end{equation}
with $\Theta_{m}=\theta_{m}/\sqrt{2}$.

In figure~\ref{fig:alignVSsequential_time} we report the energy stored in the QB as a function of time (red curve), given (assuming again $\omega_{0}$ as the reference energy) by
\begin{equation}
    E^{(3)}_{sim}(t)=\frac{\Delta}{2}\sin^{2}\Theta(t)+\frac{\Delta_{max}}{4}\left[1-\cos\Theta(t)\right]^2.
    \label{E_sim}
\end{equation}
As expected, the complete charging of the QB can be obtained here in a unique step as long as $\Theta_{m}=\pi$. Notice that a similar form of the stored energy can be obtain under proper conditions within the adiabatic approximation (see~\ref{AppC} for more details). 

Assuming, in analogy with what done for the qubit QB, the charging time as the one required to reach $E^{(3)}_{thr}=0.95\Delta_{max}$, one has also in this case $t_{c}=0.59t_m$. Notice that, for a fix $t_m$, this leads to a faster charging (greater charging power) with respect to the sequential case. 

\begin{figure}[h]
    \centering
    \includegraphics[width=0.6\textwidth]{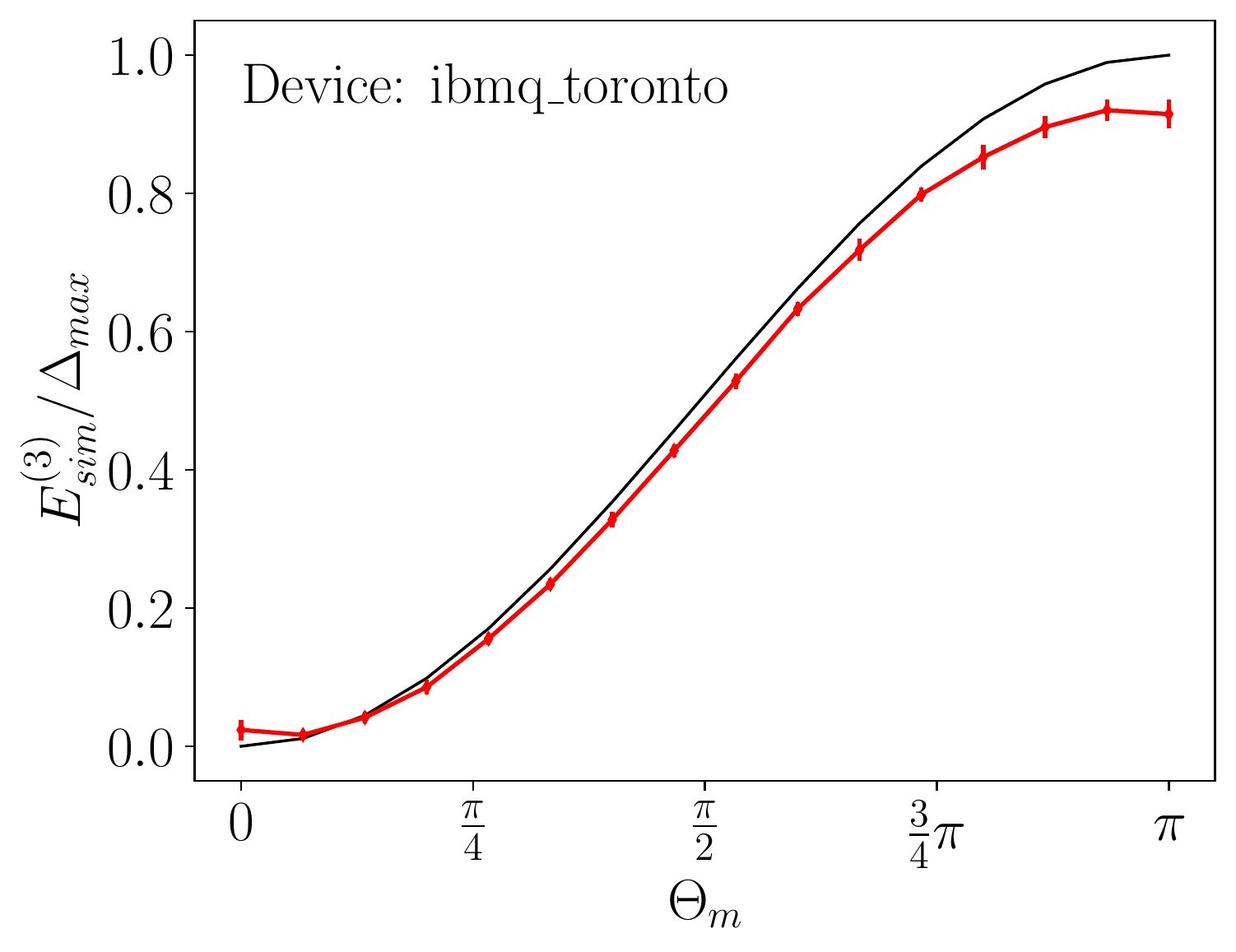}
    \caption{Energy stored in the QB (in units of $\Delta_{max}$) as a function of $\Theta_{m}$ following the simultaneous charging protocol. The black line is obtained analytically from~(\ref{E_sim}). The red points correspond to experimental data, obtained from the \emph{ibmq\_toronto} machine. We have considered the same Gaussian pulse as in the qubit case with $t_m=30\,\, \mathrm{ns}$.}
    \label{fig:align_medie30ns}
\end{figure}

The relative experimental data as a function of $\Theta_{m}$, obtained using the \emph{ibmq\_toronto} device (a machine composed by $27$ transmon circuits, of which we address the number $16$ that is characterized by the best
compromise between the longer relaxation and dephasing times and the smaller read-
out error) are reported in figure~\ref{fig:align_medie30ns} \footnote{This device has been officially retired by IBM on April 10th 2023, during our investigation.}. In is worth to note that this simultaneous charging protocol cannot be implemented on all IBM quantum machines that can be accessed via qiskit-pulse due to software constraints~\cite{IBM_private}. Also in this case data are obtained following the calibration procedure described in~\ref{sec:calibration}. The maximum energy reached is $92.0\%$ of $\Delta_{max}$. This indicates that efficiencies of the two considered protocols are very closed. Moreover, the charging occurs in roughly the same amount of time with respect to the other ($t_{c}\approx\,\,20 \mathrm{ns}$) with an analogous relaxation time ($t_{c}\approx 100\,\,\mu\mathrm{s}$). As far as we know this is the faster stable charging process reported so far for a multi-level QB. Shorter times seem out of reach in the currently available IBM quantum devices due to discretization of the signal implemented at the level of software~\cite{Gemme22}. The amount of energy stored in the qutrit in this case in almost identical to the one reported for \emph{ibm\_auckland} ($\Delta_{max}\approx 39.3\,\,\mu\mathrm{eV}$). It is worth to mention the fact that the departure from the theoretical curve could be related to errors at the level of the initialization, to discretization of pulses or to read-out errors. 

\section{Conclusions}
We have considered two experimentally relevant cases in which the dynamics of a three-level quantum battery can be treated analytically. Starting from an analysis carried out for the simpler two-level case, we have determined the charging time for: i) a sequential charging protocol where the qutrit is charged according to the two subsequent steps $|0\rangle \rightarrow |1\rangle$ and $|1\rangle \rightarrow |2\rangle$ and ii) a simultaneous charging protocol where it is possible to achieve a direct $|0\rangle \rightarrow |2\rangle$ transition. We underline the fact that the reported results for both the charging protocols are robust against cross-talks among the various circuits composing the considered machines.

We have also tested these protocols on IBM quantum devices estimating a charging time $t_{c}\approx 20\div 25\,\,\mathrm{ns}$. These times are an order of magnitude shorter with respect to a previous analysis carried out in~\cite{Hu22}, in presence of a comparable stored energy and for devices characterized by longer relaxation and dephasing times. As far as we know, these results, in particular for what it concerns the simultaneous charging, represent the fastest stable charging reported so far in the framework of multi-level solid state quantum batteries based on superconducting circuits. 

As an interesting by-product of our analysis, we have shed new light on the time dependent control of multi-level quantum systems with relevant impact in the field of quantum computation. Indeed, the possibility to use quantum devices both as qubit and as qutrit~\cite{Lierta22} or more generally qudit~\cite{Jankovic23} could make the current quantum computers more versatile broadening the panorama of future possible applications~\cite{Nguyen22}.

\ack  We would like to thank G. M. Andolina, G. Benenti, N. Bronn, F. Campaioli, N. Earnest-Noble and J. Quach for useful discussions. D. F. would like to acknowledge the contribution of the European Union-NextGenerationEU
through the “Quantum Busses for Coherent Energy Transfer” (QUBERT) project, in the framework of the
Curiosity Driven 2021 initiative of the University of Genova and through the ”Solid State Quantum Batteries: Characterization and Optimization" (SoS-QuBa) project, in the framework of the PRIN 2022 initiative of the Italian Ministry of University (MUR) for the National Research Program (PNR).
M. G. and S. V. are supported by CERN through CERN Quantum Technology Initiative. Access to IBM devices has been granted via CERN HUB. The views expressed are those of the authors, and do not reflect the official policy or position of IBM or the IBM Quantum team.

\appendix
\section{Theoretical description of the transmon qubit} \label{AppA}
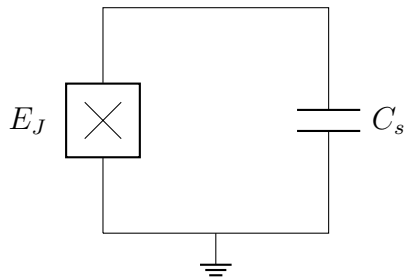
\begin{figure}
    \centering
   \begin{circuitikz} \draw
(0,0) -- (0,1)
(0,1.5) node[mixer, box only](m){} (0,2) --(0,3) -- (3,3) to[C] (3,0) -- (1.5,0)
  node[ground]{} -- (0,0)
(-1,1.5) node{$E_J$}
(3.8,1.5) node{$C_s$}
;
\end{circuitikz}
    \caption{Scheme of a superconducting circuit composed by a Josephson junction (crossed square symbols) with energy $E_J$ and a capacitance
$C_s$.}
    \label{fig:circuit_transmon}
\end{figure}

We want to provide here a simple circuital scheme leading to the two- and three-level quantum devices discussed in the main text (see figure~\ref{fig:circuit_transmon}). 

Its Hamiltonian is given by~\cite{Koch07, Krantz19}
\begin{equation}
H=4 E_{C}N^{2}-\frac{E_{J}}{2}\cos{\Phi}
\end{equation}
with 
\begin{equation}
E_{C}=\frac{e^{2}}{C_{s}}>0 
\end{equation} 
the charging energy associated to the capacitive part of the circuit ($e$ here is the charging energy and $C_{s}$ the capacitance), $E_{J}>0$ the energy associated to the Josephson junction, $N$ the Cooper pair number operator and $\Phi$ the conjugate phase operator which satisfies
\begin{equation}
\left[\Phi, N\right]=i.
\end{equation}
In the transmon limit $E_{C}\ll E_{J}$ this problem maps into the one of a particle with very small kinetic energy trapped into a cosine-like potential. Under these conditions it is possible to Taylor expand the cosine term up to the forth order obtaining 
\begin{equation}
H\approx4 E_{C}N^{2}+\frac{E_{C}}{2}\Phi^{2}-\frac{E_{C}}{24}\Phi^{4}.
\end{equation}
The previous Hamiltonian can be quantized introducing ladder operators satisfying
\begin{equation}
\left[b, b^{\dagger}\right]=1
\end{equation}
and such that 
\begin{eqnarray}
N&=& i \left(\frac{E_{J}}{32 E_{C}} \right)^{\frac{1}{4}}\left(b^{\dagger}-b \right)\\
\Phi&=& \left(\frac{2E_{C}}{E_{J}} \right)^{\frac{1}{4}}\left(b^{\dagger}+b \right).
\end{eqnarray}
According to this, one obtains an anharmonic oscillator of the Duffing type 
\begin{equation}
H\approx \omega_{P}b^{\dagger}b- \frac{E_{C}}{12}\left(b^{\dagger}+b\right)^{4}
\end{equation}
with 
\begin{equation}
\omega_{P}=\sqrt{8 E_{C} E_{J}}
\end{equation}
the so called plasma frequency of the circuit. 

In the considered limit the energy levels are very well determined already at the first order in perturbation theory, leading (up to a constant) to 
\begin{equation}
\omega_{n}=\left(\omega_{P}-E_{C}\right)n-\frac{1}{2}E_{C}n\left(n-1\right).
\end{equation}
From this we finally derive
\begin{eqnarray}
\Delta&=&\omega_{1}-\omega_{0}= \omega_{P}-E_{C}\\
\Delta'&=&\omega_{2}-\omega_{1}= \omega_{P}-2E_{C}=\Delta-E_{C}\\
\Delta_{max}&=& \omega_{2}-\omega_{0}=\Delta+\Delta'=2\omega_{P}-3E_{C}.
\end{eqnarray}
These parameters are the ones considered in the main text.
 
\section{Calibration and data analysis}
\label{sec:calibration}
The reconstruction of the state of a transmon,  after the application of a time dependent external drive, is done through a readout in the so-called dispersive regime. Here, a harmonic oscillator ($LC$ circuit playing the role of a resonator) is weakly coupled to the transmon and off resonant with respected to it~\cite{Krantz19}. In this regime the frequency of the oscillator depends of the state of the transmon. This allows for a so called non-destructive measurement~\cite{Jeffrey14} based on the fact that a monochromatic microwave with frequency $\Omega_{0}$ applied to the
resonator is modified in such a way that
\begin{equation}
\cos{\Omega_{0} t} \rightarrow A \cos(\Omega_{0} t + \chi), 
\end{equation}
with $A$ and $\chi$ real numbers representing an amplitude and a phase respectively.  
Taking into account the complex representation of the transmitted wave at a given time, one can write
\begin{equation}
Ae^{i \chi} = I + iQ,
\end{equation}
with $I$ and $Q$ real numbers. Every measurement of the transmon state is therefore reported as a point in the $(I,Q)$ plane. In order to accumulate proper statistics, the machine performs multiple runs ($1024$ in default settings). They are typically very scattered, requiring a further analysis to extract meaningful information from them. 
We have classified the points according to the three relevant states of the system ($|0\rangle$, $|1\rangle$ and $|2\rangle$) by means of scikit-learn, an open source machine learning library based on the Python programming language \cite{scikit-learn}. We have used support vector machines method with linear kernel function. This method takes as input two arrays: an array $X$ of shape $(n\_samples, \,n\_features)$ holding the training samples, and an array $y$ of class labels, that can be strings or integers, of shape $(n\_samples)$. In our case we have $3072$ samples with 2 features, $I$ and $Q$ and $3$ possible labels, $0$, $1$ or $2$.  After being trained, the model can be used to classify new values.

\begin{figure}[h]
    \centering
    \includegraphics[width=0.6\textwidth]{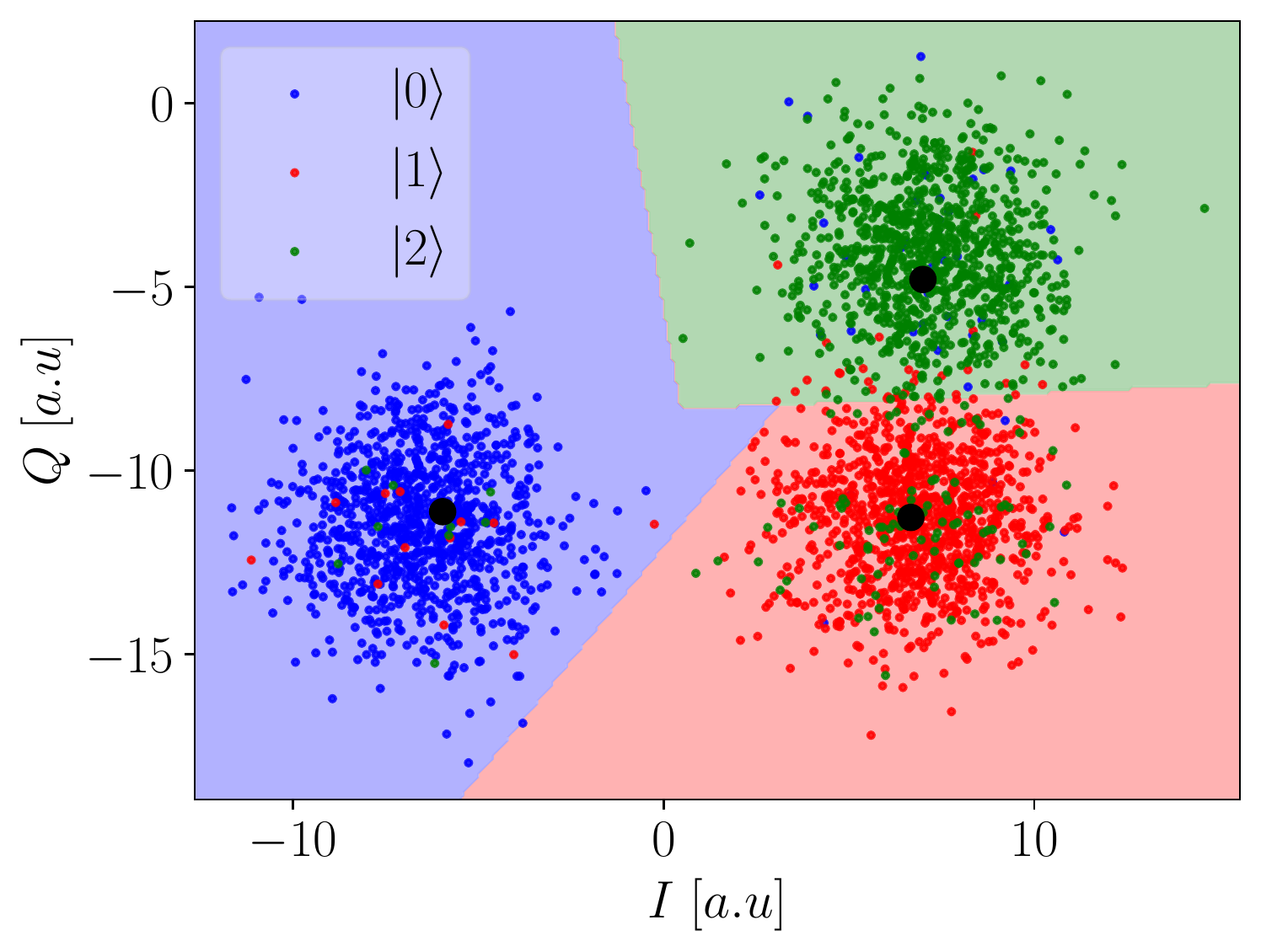}
    \caption{Example of data distribution associated to the measurements of the state $\ket{0}$ (blue dots), $\ket{1}$ (red dots) and $\ket{2}$ (green dots) in the $(I,Q)$ plane (in arbitrary units) for the \emph{ibm\_auckland} device. Big black dots indicate the centers of the different distributions, while straight lines separates the regions associated to every state. The efficiency of the considered separation is roughly $95.5\%$ for the ground state, $95.7\%$ for the first excited state and $90.0\%$ for the second excited state. For each state shown in the plot, we have considered $1024$ runs ($3072$ in total).}
    \label{fig:divisione_qutrit}
\end{figure}

Figure~\ref{fig:divisione_qutrit} shows an example of data distribution with the colored regions graphically representing the data labels. According to this picture, the energy stored in a qutrit QB (with respect to the ground state), can be determined through the relation
\begin{equation}
E(\eta)=\Delta\mathcal{P}_1(\eta)+\Delta_{max}\mathcal{P}_2(\eta)
\end{equation}
with 
\begin{equation}
\mathcal{P}_i(\eta)=|\braket{\Psi(\eta)|i}|^2
\end{equation}
and $\eta=\varphi_{m}, \Theta_{m}$ depending on the considered charging protocol. 

\begin{figure}[h]
    \centering
    \includegraphics[width=1\textwidth ]{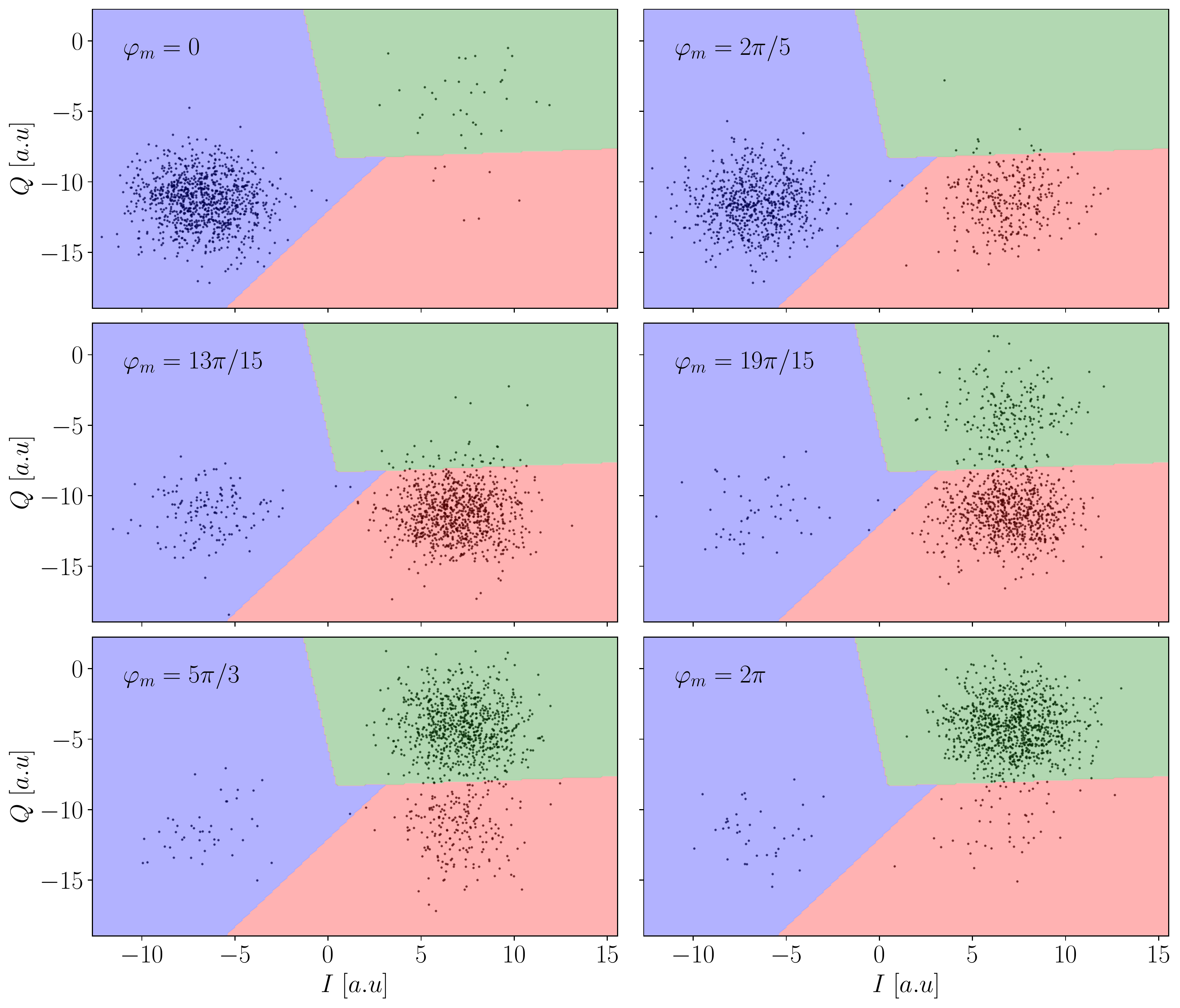}
    \caption{Example of data distribution associated to the sequential charging protocol. Each plot shows the results (black dots) in the $(I,Q)$ plane (in arbitrary units) as a function of $\varphi_{m}$. For each state shown in the plot, we have considered $1024$ runs. These measurements have been carried out using the \emph{ibm\_auckland} device. The background is coloured according to what discussed in the calibration phase. In particular the blue part is classified as $\ket{0}$, the red one as $\ket{1}$ and the green one as $\ket{2}$.}
    \label{fig:evoluzione_seq}
\end{figure}

\begin{figure}[h]
    \centering
    \includegraphics[width=1\textwidth ]{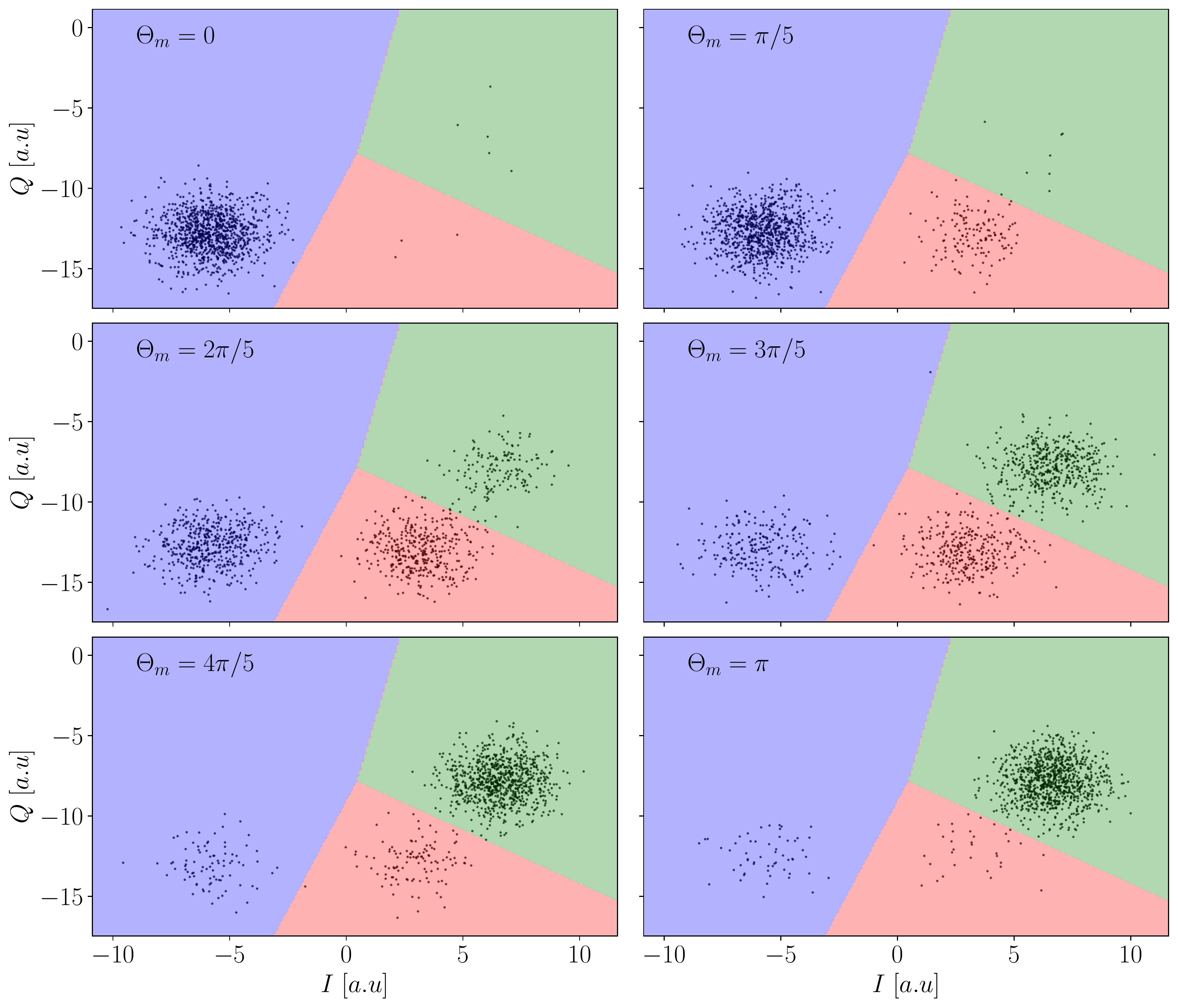}
    \caption{Example of data distribution associated to the simultaneous charging protocol. Each plot shows the results (black dots) in the $(I,Q)$ plane (in arbitrary units) as a function of $\Theta_{m}$. For each state shown in the plot, we have considered $1024$ runs. These measurements have been carried out using the \emph{ibmq\_toronto} device. The background is coloured according to what discussed in the calibration phase. In particular the blue part is classified as $\ket{0}$, the red one as $\ket{1}$ and the green one as $\ket{2}$.}
    \label{fig:evoluzione_align}
\end{figure}

In figures~(\ref{fig:evoluzione_seq}) and (\ref{fig:evoluzione_align}) we show the different evolution of the state of the qutrit in the $(I,Q)$ plane considering the sequential and simultaneous charging respectively. In particular, while in the former case there is an intermediate situation in which the system is in the state $|1\rangle$ with high probability, this doesn't happens in the latter. 

\section{Adiabatic charging of the three-level QB} \label{AppC}

An alternative way to charge the qutrit QB realizing a stable $|0\rangle\rightarrow |2\rangle$ transition involves a classical charging (see~(\ref{Charging_3})), with two identical time dependent drives such that 
\begin{equation}
f_{1}(t)=f_{2}(t)=f(t)
\end{equation}
and 
\begin{equation}
\Omega_{1}=\Omega_{2}=\frac{\Delta+\Delta'}{2},
\end{equation}
namely resonant with a unique frequency given by the average of the two level spacing~\cite{Willsch23}. Under such conditions, assuming again the RWA, one obtains the new effective Hamiltonian 
\begin{equation}
 \hat{\mathcal{H}}_{eff}^{(3)}(t)= 
    \frac{g}{2}f(t)\left( \matrix{
        0 & e^{-i \delta t} & 0\cr
        e^{i \delta t} & 0 & e^{i \delta t}\cr
        0 & e^{-i \delta t} & 0
    \cr} \right),
    \label{Charging_delta}
\end{equation}
with 
\begin{equation}
\delta=\frac{\Delta-\Delta'}{2}.
\end{equation}
Notice that, according to the derivation reported in~\ref{AppA}, this parameter is positive and can be written only in term of the transmon charging energy, namely  
\begin{equation}
\delta=\frac{E_{C}}{2}.
\end{equation}

In order to solve the dynamics, in this case it is possible to consider a full numerical approach. However, in the following we will proceed along a different path working on the the abiabatic approximation~\cite{Berry84}. This will allow us to have a better insight of the physics of the system. In this case the state of the system at a given time $t$ can be approximated as 
\begin{equation}
|\Psi'(t)\rangle\approx\sum_{\sigma}c_{\sigma} e^{-i \int^{t}_{0}E_{\sigma}(\tau) d\tau} e^{i \gamma_{\sigma}(t)}|\Psi_{\sigma}(t)\rangle\qquad \sigma=B, \pm.
\label{psi_adiabatic}
\end{equation}

In the above expression one needs to take into account the instantaneous eigenstates of the Hamiltonian in~(\ref{Charging_delta})
\begin{equation}
    |\Psi_{B}(t)\rangle=\left( \matrix{
        -\frac{1}{\sqrt{2}}\cr
        0\cr
        \frac{1}{\sqrt{2}}
    \cr} \right);\qquad |\Psi_{\pm}(t)\rangle=
    \left( \matrix{
        \frac{1}{2}\cr
        \pm \frac{e^{i \delta t}}{\sqrt{2}}\cr
        \frac{1}{2}
    \cr} \right)
\end{equation}
with instantaneous energy eigenvalues
\begin{eqnarray}
E_{B}(t)&=&0\nonumber\\ 
E_{\pm}(t)&=&\pm \frac{g}{\sqrt{2}}f(t).
\end{eqnarray}

Others important terms which compare in~(\ref{psi_adiabatic}) are the geometric or Berry phases~\cite{Berry84}
\begin{eqnarray}
\gamma_{B}(t)&=&i\int_{0}^{t}d \tau \langle \Psi_{B}(t)| \frac{d}{d\tau}|\Psi_{B}(t)\rangle=0\nonumber\\
\gamma_{\pm}(t) &=&i\int_{0}^{t}d \tau \langle \Psi_{\pm}(t)| \frac{d}{d\tau}|\Psi_{\pm}(t)\rangle=- \frac{\delta t}{2}.
\end{eqnarray}

Taking into account the initial condition already discussed in the main text, namely 
\begin{equation}
c_{B}=\frac{1}{\sqrt{2}}; \qquad c_{\pm}=\frac{1}{2}, 
\end{equation}
one finally obtains 
\begin{equation}
    \ket{\Psi'(t)}\approx\left( \matrix{
    \frac{1}{2}\left[\cos\Theta(t)e^{-i\frac{ \delta t}{2}}-1\right]\cr
    -\frac{i}{\sqrt{2}}\sin\Theta(t)e^{ i\frac{\delta t}{2}}\cr
    \frac{1}{2}\left[\cos\Theta(t)e^{-i\frac{ \delta t}{2}}+1\right]
    \cr} \right),
\end{equation}
with stored energy 
\begin{equation}
    E^{(3)}_{ad}(t)\approx\frac{\Delta}{2}\sin^{2}\Theta(t)+\frac{\Delta_{max}}{4}\left[1-2\cos\Theta(t)\cos\left(\frac{\delta t}{2} \right)+\cos^{2}\Theta(t)\right].
    \label{E_ad}
\end{equation}

\begin{figure}[h]
    \centering
    \includegraphics[width=0.6\textwidth ]{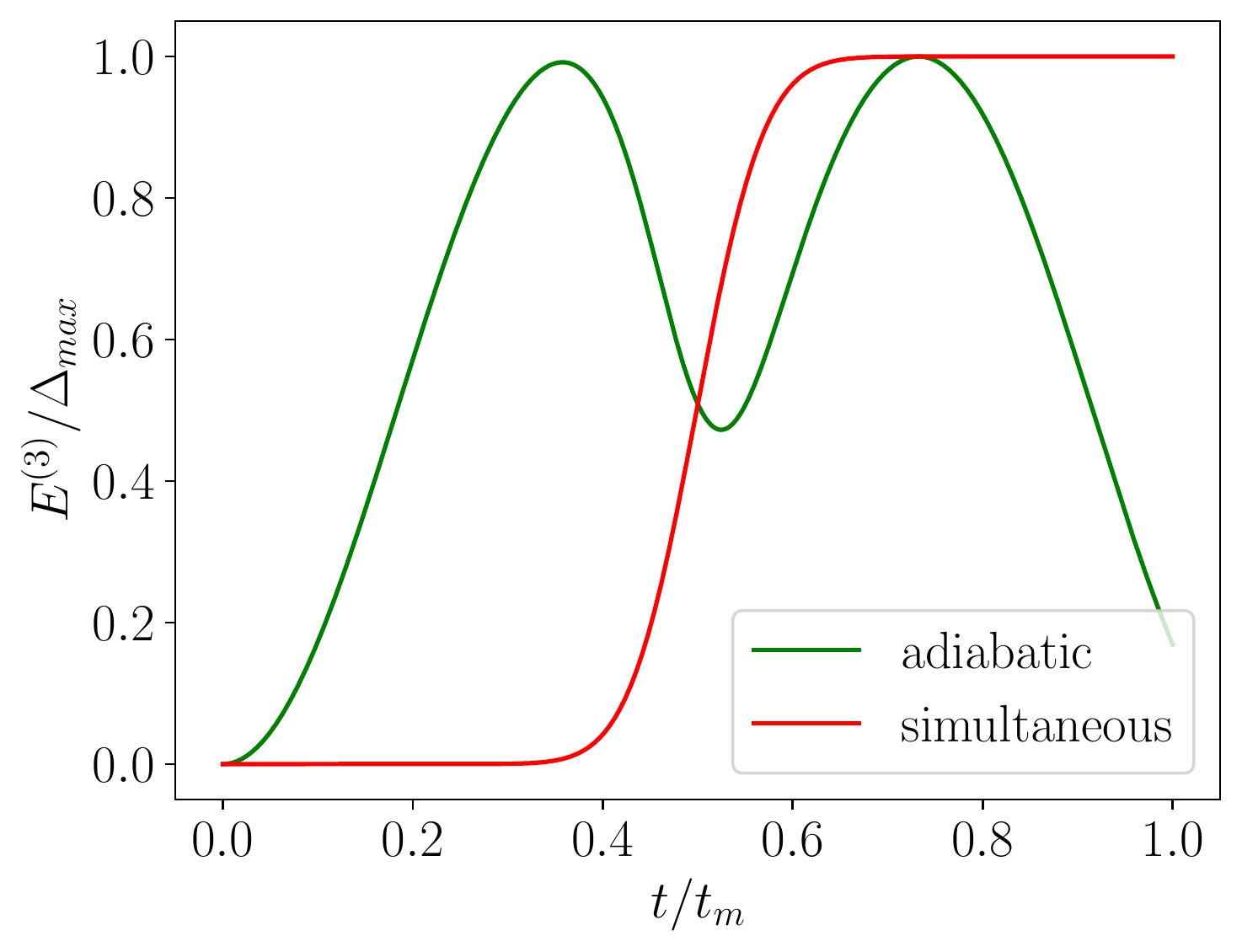}
\caption{Energy stored in the qutrit QB (in units of $\Delta_{max}$) as a function of $t$ (in units of $t_{m}$) for both an adiabatic (green curve) and simultaneous (red curve) charging protocol. Here we have considered the same Gaussian pulses as in the qubit case.}
    \label{fig:alignVSadiabatic_time}
\end{figure}

In the regime where both approaches are applicable, the adiabatic charging usually leads to a faster but less stable charging with respect to the simultaneous (see figure (\ref{fig:alignVSadiabatic_time})).

\end{document}